\documentclass{fundam}

\publyear{22}
\papernumber{2102}
\volume{185}
\issue{1}

\usepackage[
pdfauthor={Vegard Steinsland (vste@hvl.no), Lars Michael Kristensen (lmkr@hvl.no) and Shujun Zhang (shz@hvl.no)},
pdftitle={Modelling and Validation of Power Electronics Converter Systems using Coloured Petri Nets},
pdfsubject={Fundamenta Informaticae special issue - Petri Nets 2022},
pdfkeywords={Petri Net, CPN Tools, Power converter, Embedded systems, Cyber-physical systems},
pdfcreator={pdfTeX 3.141592653-2.6-1.40.24 (TeX Live 2022)},
bookmarks=true,
hidelinks
]{hyperref}

\hypersetup{bookmarksdepth=section}

\usepackage{url} % takes care of hyperlinks, preferred over hyperref
\usepackage[ruled,lined]{algorithm2e}% provides Algorithm environment
\usepackage{graphicx}% allows for inclusion of graphic files (figures)
\usepackage{tikz}
\usetikzlibrary{automata, positioning, arrows}

\usepackage{import}
\usepackage{standalone}

\usepackage{siunitx}

\usepackage{pgfplots}
\usepackage{tikz}
\pgfplotsset{compat=newest}
\usepgfplotslibrary{units,external}
\usepgfplotslibrary{fillbetween}
\usetikzlibrary{shapes.geometric,graphs,shapes,arrows,positioning,calc,decorations.markings,fit,decorations.pathmorphing,matrix,chains,scopes,intersections, decorations.pathreplacing, patterns}
\usepackage{blox}
\usepackage{tikz-dimline}
\usepackage[european,siunitx,RPvoltages]{circuitikz}

\makeatletter
\def\relativepath{\import@path}
\makeatother

\usepackage{amsmath}
\usepackage{cleveref}
\usepackage{glossaries-extra}
\newabbreviation{gabr-cpn}{Colored Petri nets}{}
\newabbreviation[parent=gabr-cpn]{abr-cpn}{CPN}{Colored Petri nets}
\newabbreviation[parent=gabr-cpn]{abr-sml}{SML}{Standard Meta Language}

\newabbreviation[parent=gabr-cpn]{abr-des}{DES}{discrete event system}
\newabbreviation[parent=gabr-cpn]{abr-cps}{CPS}{cyber-physical system}

\newabbreviation[parent=gabr-cpn]{abr-stg}{STG}{Signal Transition Graphs}

\newabbreviation{gabr-gen}{General}{}
\newabbreviation[parent=gabr-gen]{abr-hw}{HW}{hardware}
\newabbreviation[parent=gabr-gen]{abr-sw}{SW}{software}

\newabbreviation[parent=gabr-gen]{abr-si}{SI}{International System of Units}

\newabbreviation[parent=gabr-gen]{abr-gui}{GUI}{graphical user interface}

\newabbreviation[parent=gabr-gen]{abr-rt}{RT}{real-time}
\newabbreviation[parent=gabr-gen]{abr-io}{IO}{input/output}

\newabbreviation{gabr-mmc}{MMC}{}
\newabbreviation[parent=gabr-mmc]{abr-mmc}{MMC}{Modular Multilevel Converter}
\newabbreviation[parent=gabr-mmc]{abr-sm}{SM}{sub-module}
\newabbreviation[parent=gabr-mmc]{abr-adc}{ADC}{analog to digital converter}
\newabbreviation[parent=gabr-mmc]{abr-pwm}{PWM}{Pulse Width Modulation}

\newabbreviation{gabr-zynq}{Zynq}{}
\newabbreviation[parent=gabr-zynq]{abr-axi}{AXI}{Advanced eXtensible Interface}
\newabbreviation[parent=gabr-zynq]{abr-soc}{SoC}{System-on-Chip}
\newabbreviation[parent=gabr-zynq]{abr-fpga}{FPGA}{field-programmable gate array}
\newabbreviation[parent=gabr-zynq]{abr-arm}{ARM}{Advanced RISC Machines}
\newabbreviation[parent=gabr-zynq]{abr-hdl}{HDL}{Hardware Description Language}
\newabbreviation[parent=gabr-zynq]{abr-ps}{PS}{processing system}
\newabbreviation[parent=gabr-zynq]{abr-pl}{PL}{programmable logic}

\glsdisablehyper

\crefname{equation}{Eq.}{Eqs.}
\Crefname{Equation}{Eq.}{Eqs.}

\crefname{figure}{Fig.}{Figs.}
\Crefname{figure}{Fig.}{Figs.}

\crefname{section}{Sect.}{Sects.}
\Crefname{section}{Sect.}{Sects.}

\crefname{table}{Table}{Table}
\Crefname{table}{Table}{Table}

\crefname{lstlisting}{Listing}{Listings}
\Crefname{lstlisting}{Listing}{Listings}

\usepackage{subfig}
\usepackage{float}

\usepackage{tabularx}
\usepackage{multirow}

\usepackage{listings}
\usepackage{xcolor}
\definecolor{eclipseBlue}{RGB}{42,0.0,255}
\definecolor{eclipseGreen}{RGB}{63,127,95}

\usepackage{setspace}

\newcommand{\cpntxt}[1]{\textsf{#1}} % text entry linking CPN figure

\newcommand{\smltxt}[1]{{\ttfamily#1}} % text entry linking code listings

\usetikzlibrary{external}
\begin{document}

\title{Modelling and Validation of Power Electronics Converter Systems using Coloured Petri Nets}

\author{Vegard Steinsland \\
Department of Computer science, Electrical engineering and Mathematical sciences \\
Western Norway University of Applied Sciences
\and Lars Michael Kristensen \\
Department of Computer science, Electrical engineering and Mathematical sciences \\
Western Norway University of Applied Sciences
%\and Reza Arghandeh\\
%Department of Computer science, Electrical engineering and Mathematical sciences \\
%Western Norway University of Applied Sciences\\ Inndalsveien 28, 5063 Bergen, Norway\\
\and Shujun Zhang \\
Department of Computer science, Electrical engineering and Mathematical sciences \\
Western Norway University of Applied Sciences\\
}

\address{V. Steinsland: \texttt{vegard.steinsland{@}hvl.no}}

\maketitle

\runninghead{Steinsland et. al}{Design and Validation of Power Electronics Converters using CPNs}
\vspace{-7em}

\begin{abstract}
We apply Coloured Petri Nets (CPNs) and the CPN Tools to develop a formal model of an embedded system consisting of a power converter and an associated controller. Matlab/Simulink is the de-facto tool for embedded control and system design, but it relies on informal semantics and has limited support for transparent and integrated specification and validation of both the power converter electronics, controller (hardware), and the control logic (software). The contribution of this paper is to develop a timed hierarchical CPN model that mitigates the shortcomings of Simulink by relying on a Petri net formalisation. We demonstrate the application of our approach by developing a fully integrated model of a buck power converter with controller in CPN Tools. Furthermore, we perform time-domain simulation to verify the capability of the controller to serve the control objectives. To validate the developed CPN model, we compare the simulation results obtained in an open-loop configuration with a corresponding implementation in Simulink.
%\lmk{Update open/close-loop description and revise wrt. augmented experimental evaluation} 
The experimental results show correspondence between the CPN model and the Simulink model. As our CPN model reflects the fully integrated system, we are able to compare CPN simulation results to measurements obtained with a corresponding implementation in real hardware/software and compare closed-loop with open-loop configuration. The results show alignment for the steady state while further refinement of the control algorithm and validation is required.
\end{abstract}

\begin{keywords}
	Formal methods, power electronics, Coloured Petri nets, cyber-physical systems
\end{keywords}

\section{Introduction}

The development of power electronics converter systems is a multi-disciplinary task where the immediate time-critical response is related to the integrated system of electrical, electronics, cybernetics and embedded control design with software forming a \gls{abr-cps} \cite{wisniewskiDesignVerificationCyberPhysical2019}. Formal modelling is a valuable tool for understanding power electronics converters system and to control their behaviour, and can aid the design process from initial evaluation of proposed circuit topologies through physical prototyping with integration of the controller, to iterative improvements in both the physical design and the control. Given the comprehensive challenges in the design phase, abstraction and simplification of the system parts are needed to obtain models at the appropriate abstraction level to achieve and validate the design objective(s).

In power electronics and embedded control system design, Matlab/Simulink tool \cite{szczesniakOverviewControlAlgorithm2021} is one of the de-facto standards and is supported by an integrated tool chain. Within the tool chain of MathWorks, the Simulink tool provides predefined modelling objects with approximation of the physical systems (via differential equation descriptions), control objects, tools for embedded code generation, and tools for real-time \gls{abr-hw} monitoring to support integrated design and rapid prototyping. However, Matlab/Simulink relies on informal semantics and several researchers have identified application shortcomings implied by the lack of tools for model checking and validation
\cite{barbotIntegratingSimulinkModels2018,beraRelationshipSimulinkPetri2014,gutierrezRealtimeEmulationBoost2019,singhSpecificationGuidedAutomatedDebugging2020,wisniewskiPetriNetBasedSpecification2019}.

The lack of formal semantics in Matlab/Simulink also implies challenges when executing embedded code on real hardware, where the actual performance has timing properties relying on the integrated \gls{abr-cps}. This makes it difficult to validate the integrated system performance based on other metrics than the physical response.
This means that the approach is partly non-transparent and the designer is somehow blindfolded by the complex power electronics converter system behaviour due to the lack of tools in Matlab/Simulink to perform formal validation of the \gls{abr-cps}.
For simple power electronics converters such as the buck converter, the lack of formalization may not be a practical challenge. However, when developing systems with higher complexity such as a \gls{abr-mmc} and systems of converters forming microgrids, the lack of formal semantics and transparency may lead to unexpected events in the control schemes as discussed in  \cite{mboupPetriNetsControl2009,ortizNovelStrategyDynamic2020,wisniewskiDesignVerificationCyberPhysical2019}.

The contribution of this paper is to explore the application of Petri nets to address the shortcomings of Matlab/Simulink discussed above. Our aim is to provide an integrated modelling approach that provides transparency and traceability on both the hardware part and on the control logic part of the CPS.
Being based on Petri nets, our approach provides a formal semantics that can serve as a basis for model validation when applied for the design and implementation of power electronics converter systems. We rely on high-level Petri nets in order to conveniently model data related to sensed values (physical response), calculations, signalling, and system parameters. We specifically apply Coloured Petri Nets (CPNs) \cite{jensenColouredPetriNets1997} and CPN Tools \cite{jensenColouredPetriNets2007} as it supports the modelling and simulation of time, and the construction of hierarchical and parametric models divided into several modules. The latter is important in order to ensure transparency and scalability to larger power electronics converter systems.

As a case study, we focus on modelling the dynamics of a buck power electronics converter integrated with an embedded controller, and we use simulation and physical models to evaluate and compare the system performance in the time-domain.
First the CPN model of the buck converter dynamical response is compared with the Matlab/Simulink implementation with fixed point operation, thus reducing the complexity. This is followed by results obtained via the integrated CPN modelling of the buck converter and the controller aimed at assessing the objective as an integrated \gls{abr-cps} with closed-loop control logic serving the control objective of fixed output voltage. To further validate the constructed CPN model, we validate the CPN simulation against measurements obtained with a corresponding implementation of the power converter system in real hardware. 
%\lmk{Could be a need for updating this section given the augmented experimental evaluation}

The rest of this paper is organised as follows. \Cref{sec:cpn-implementation-cps} gives an overview of our modelling approach and the power converter system comprised of the buck converter topology description and the controller. \Cref{subsec-cpn-buck} concentrates on the modelling of the buck converter, and \cref{subsec:controller-interface} presents the modelling of the embedded controller. \Cref{sec-simulink} provides a brief description of the Matlab/Simulink model that we use for comparison. The experimental results are presented in \Cref{sec-results}. In \cref{sec-sum-conc}, we present our conclusions, discuss related work, and suggest directions for future work to further develop our modelling framework in the domain of power electronics systems. We assume that the reader is familiar with the basic concepts of Petri Nets. We informally introduce the concepts of Hierarchical Timed Coloured Petri Nets (High-level Petri nets) as we proceed with the presentation of our formal modelling. The reader interested in the full and formal definition of the syntax and semantics of Coloured Petri Nets is referred to \cite{cpnbook2009}.

\tikzexternalenable

\section{Modelling Approach and Overview of the CPN Model}
\label{sec:cpn-implementation-cps}

The developed CPN model comprises the complete system and consists of the buck converter integrated with the controller. Our modelling approach allows for system parametrisation of the physical converter model, the system interfaces, and the controller properties.
The objective of the integrated buck power converter system is to reduce the input voltage ($V_i$) to a given output voltage ($V_o$) based on a \gls{abr-pwm} control signal from the controller to the converter.

\begin{figure}[h]
	\centering
	\scalebox{0.55}{
%		\subimport{}{figs/physical-buck-zynq-scope-marked}
		\includegraphics
		[clip, trim=1cm 0.5cm 1.5cm 0.5cm]
%		%trim=left botm right top
		{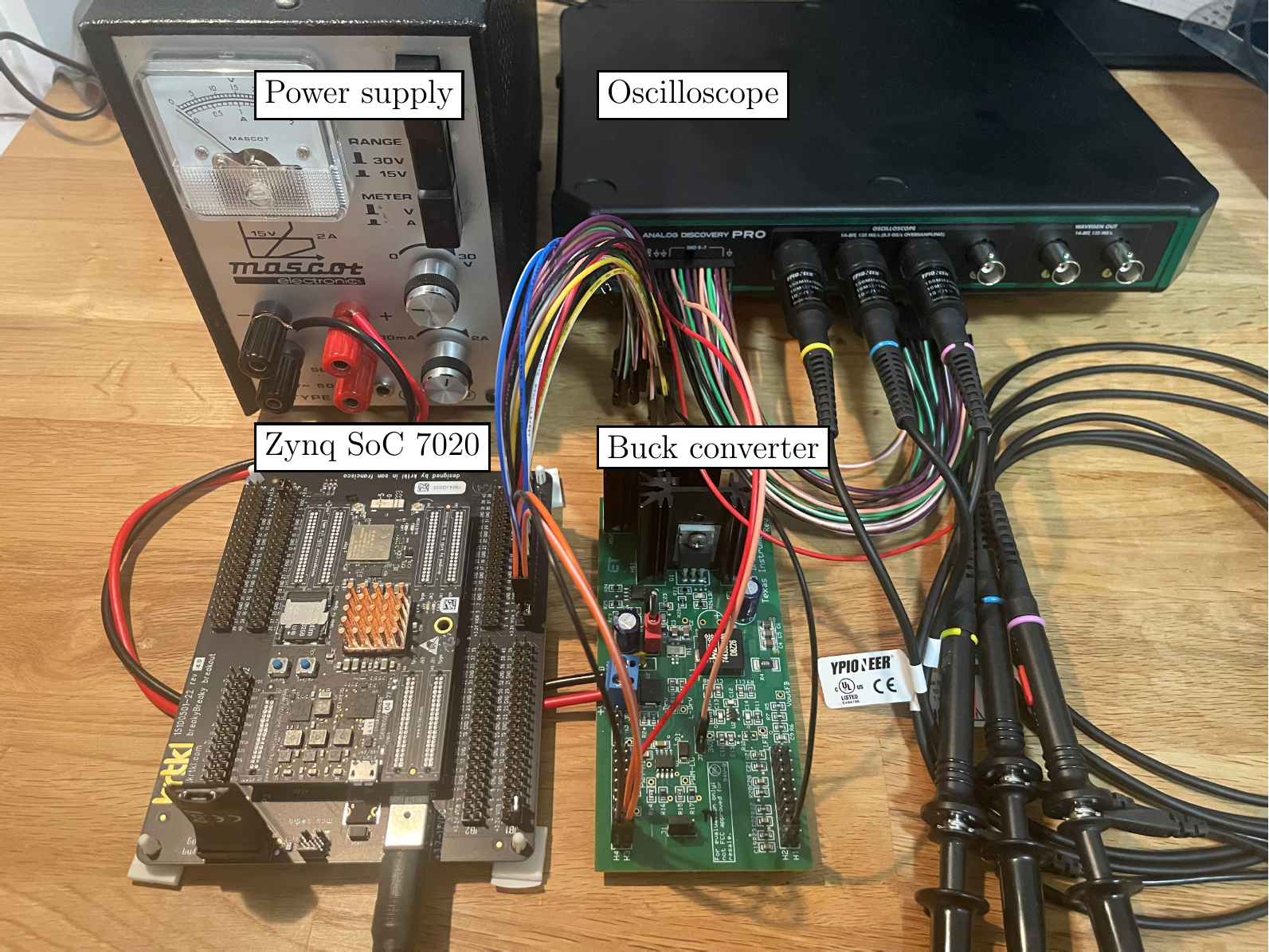}
		}
	\caption{The modelled CPS with KRTKL Snickerdoodle Black Zynq SoC 7020 controller and buck-converter. Connected to laboratory power supply and oscilloscope.}
	\label{fig:physical-buck-zynq-scope-temp}
\end{figure}

\Cref{fig:physical-buck-zynq-scope-temp} shows the actucal physical CPS being modelled which is also the system against which we validate the simulation results.
The controller is an embedded Xilinx Zynq \gls{abr-soc} 7020 hardware controller that consists of both an \gls{abr-arm} \gls{abr-ps} and a \gls{abr-fpga} \gls{abr-pl} in one single package.
The main advantages of \gls{abr-soc} with \gls{abr-fpga} are the parallel task execution feature, very high processing speed, and the integrated \gls{abr-arm} processor.
These attractive features has resulted in the growing popularity of \gls{abr-soc} being used in embedded controls of power electronics \cite{ormaetxeaMatrixConverterProtection2011,wisniewskiPetriNetBasedSpecification2019}.

\begin{figure}[ht]
	\centering
	\scalebox{0.35}{\includegraphics
		{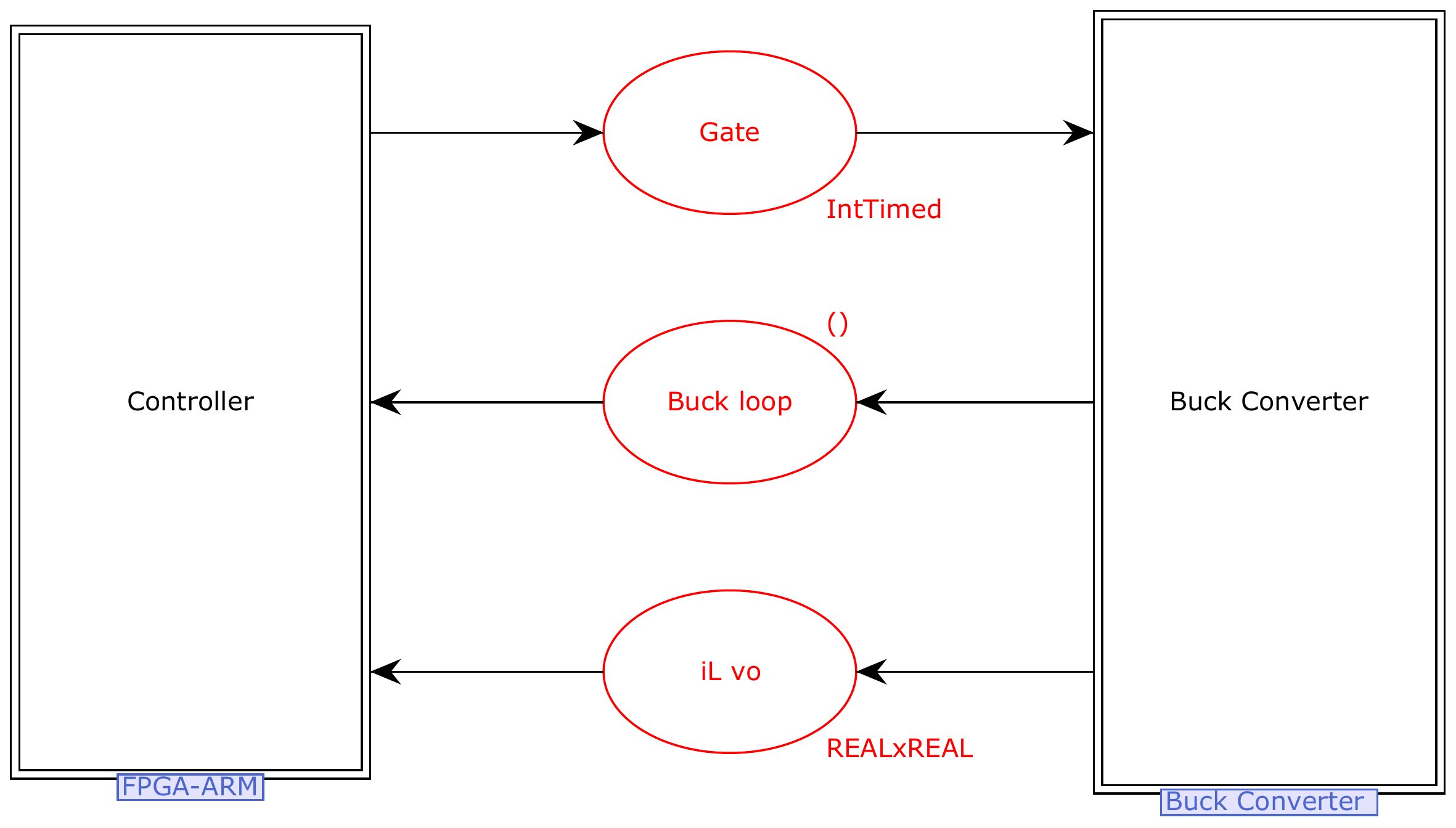}}
	\caption{Top-level CPN module of the CPS with integrated modelling of the controller (left), buck converter (right), and their interfaces (middle).}
	\label{fig:cpn-zynq-buck-full}
\end{figure}

The CPN model developed for the \gls{abr-cps} is hierarchically organised into seven modules with the top-level module shown in \cref{fig:cpn-zynq-buck-full}.  Hierarchical CPN models based on the concept of substitution transitions that have associate submodules. Substitution transitions constitute a syntactical structuring mechanism used to represent system components and compound behaviour.  The detailed behaviour represented by a substitution transition is then captured by the associated submodule which may again contain substitution transitions. Substitution transitions are by convention drawn as a rectangle with a double-lined border. The basic principle of of interfacing a substitution transition to the associated submodule is to link socket places connected to the substitution transition with port places on the associated submodule. Any tokens added/removed to the port place with then also be added/removed from the socket place and vice versa. Linking a socket and a port place hence creates a new compound place and make the port place serve as an interface for the submodule.

 The substitution transition \cpntxt{Buck Converter} in \cref{fig:cpn-zynq-buck-full} represents the physical electrical converter system and the substitution transition \cpntxt{Controller} represents the hardware and the control logic of the controller. The rectangular tags at the bottom of each of the two substitution transitions is the name of the associated submodule. The detailed CPN modelling of the buck converter and the controller will be presented in \cref{subsec-cpn-buck} and \cref{subsec:controller-interface}, respectively. The socket places \cpntxt{Gate} and \cpntxt{Buck loop} connected to the substitution transitions are used to pass the control signal to the converter and receive the acknowledgment of the converter state, respectively. The converter inductor current ($i_L$) and output voltage ($v_o$) measurements are interfaced with the controller and modelled by the socket place named \cpntxt{iL vo}.

 The colour sets (types) determining the kind of tokens that can reside on the places is provided below each place and defined in \cref{lst:cpn-top-level}. There is no colour set provided for the place \cpntxt{Buck loop} as it has the default colour set \cpntxt{UNIT} containing the single value \smltxt{()} (unit). This place initially contains a single ()-token as specified by the initial marking to the upper right of the place.

%\begingroup

\vspace*{1em}
\begin{lstlisting}[caption={Colour set definitions used in the top-level module.},label={lst:cpn-top-level}]
	colset IntTimed  = int timed;

	colset iL        = real;
	colset vo        = real;
	colset REALxREAL = product iL * vo;
\end{lstlisting}
%\endgroup

 The CPN model of the \gls{abr-cps} is a timed CPN model as we need to capture the timing aspects of the control-loop in order to properly reflect the modelled system. In particular, it can be seen in \cref{lst:cpn-top-level}. that the colour set \smltxt{IntTimed} is a timed colour set which means that tokens from this colour set will carry a time stamp.

 The time concept of CPN models is a based on a discrete global model clock and associating time stamps to the tokens. The time stamps of a token specifies the earliest time at which the tokens can be consumed by a transition. When no transitions are enabled at the current model time, time is advanced to the earliest next time at which one or more transitions becomes enabled.  The buck converter is a continuous-time dynamic system that evolves over time and the model description is implemented as a \gls{abr-des} together with the digital controller with real-time properties. For the CPN modelling of the CPS as a real-time system, we must therefore consider the continues-time dynamics of the modelled artifact and apply appropriate time discretisation to achieve the desired modelling accuracy.

The physical behaviour of the buck converter over time is described through differential equations which we have implemented in CPN Tools with emphasis to clearly show graphically how each step of the calculation is performed (see \cref{subsec-cpn-buck}). The reason for selecting this approach is to provide traceability in how the mathematical description is implemented to the explicit case of the buck converter. By default, the simulated timed properties in CPN Tools have no intrinsic relationship to real-time. To align the CPN Tools simulation time with the real-time physical properties, we introduce the simulated time iteration constant ($k$) and the discrete time-step ($T_s$) in the discretisation of both the buck converter and controller. It is the system under investigation that governs the required refinement of the simulation model discrete time-step ($T_s$).
The buck converter holds switching states and therefore the discrete time-step ($T_s$) must be smaller than the cyclic period ($T_p$) to represent the dynamic response in the electric circuit in-between a switching event.
The discrete time-step ($T_s$) by the simulator must be less than ($T_p$) to represent the circuit dynamics correctly.
A drawback of making the discrete time-step ($T_s$) too small is its negative impact on the performance of the CPN Tools simulation engine. Analytical tools and domain-specific experience/experimental results may provide insights into determining a suitable discrete time-step ($T_s$).

The simulated time in the CPN model is represented via two functions: \smltxt{Dynsys()} and \smltxt{CPU\_CLK()}. The task of the \smltxt{Dynsys} function is to keep track on the time iteration of the dynamic system that forms the buck converter. As the dynamic system has the highest discretisation requirement in the simulation, the overall time reference is set as \smltxt{Dynsys} and holds the iteration relation to the real-time trough the discrete time-step ($T_s$).
The function \smltxt{CPU\_CLK} is used to describe the timing of the controller CPU cycle, and is aligning the cycle frequency of the controller to the real-time properties of \smltxt{Dynsys}.
%\lmk{Should we show the implementation of the two functions maybe as equations rather than SML code}

\section{Buck Converter CPN Modelling} \label{subsec-cpn-buck}

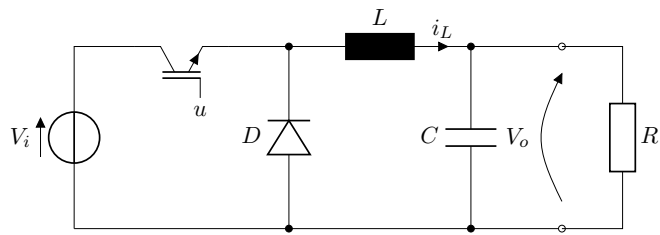
\begin{figure}[ht]
	\centering
	\tikzsetnextfilename{t-buck-converter-circuit-diagram2}
	\scalebox{0.8}{\begin{circuitikz}
	\draw (0,-4) node[Lnigbt,rotate=90](npn){}++(2,0);
	
	\draw (npn.B) node[below]{$u$};
	
	\draw (npn.C) to[short] ++(-1,0) to[european voltage source,v_<=$V_i$] ++(0,-3) coordinate(dc-gnd); %node[](dc-gnd){};
	
	\draw (npn.E) to[short] ++(1,0) coordinate(mid-top) to[L,l=$L$,invert,i=$i_L$] ++(3,0) coordinate(out-m-top) to[C,l_=$C$] ++(0,-3) coordinate(out-m-bot) to[short] ++(-3,0) coordinate(mid-bot) to[short] (dc-gnd);
	
	\draw (mid-bot) to[D,l=$D$,*-*] (mid-top);

	\draw (out-m-top) to[short,*-o] ++(1.5,0) coordinate(out-top);
	\draw (out-m-bot) to[short,*-o] ++(1.5,0) coordinate(out-bot);
	
	\draw (out-top) to[open,v<=$V_o$] (out-bot);
	
	\draw (out-top) to[short] ++(1,0) to[R,l=$R$] ++(0,-3) to[short] (out-bot);
\end{circuitikz}%
}
	\caption{Buck converter principal electrical topology.}
	\label{fig:buck}
\end{figure}

The buck converter is an electrical design topology used to step down the input voltage ($V_i$) to the desired output voltage ($V_o$). With widespread use in consumer electronics devices there is a growing demand also to utilize the buck converter in high power application in the grid, transportation electrification and renewable integration.
The principal electrical topology of the buck converter and its main components is shown in \cref{fig:buck}.
The components introducing circuit states are the non-linear semiconductor elements with the active controlled transistor ($u$) and the passively controlled diode ($D$). An semiconductor component, transistor and diodes, gives the ability to switch (turn off and on) the device and we may benefit from the circuit topology based on the switching states available through the transistor and the forced state of the diode. 
Especially, the linear inductor element ($L$) induces a voltage (electromotive force) from the flowing current, which strength is dependent on current strength and polarity dependent on current direction.
In addition, the capacitor element ($C$) act as a filter to reduce the voltage ripple on the output and over the load resistance ($R$) caused by the switching of inductor element.
All components values are derived from the \gls{abr-si}, relating properties with respect to voltage ($V$) and current ($I$), where the inductance is expressed in Henry (H), capacitance Farad (F) an resistance in Ohm ($\Omega$).
The uppercase inscription for components, currents, and voltages, is relating the average values while the lowercase inscriptions describes instantaneous values.
In most basic circuit analysis of the buck converter (ignoring losses), the steady-state fundamental description of the relation between the input voltage ($V_i$) and the output voltage ($V_o$) may be described \cite{mohanPowerElectronicsConverters1995} with a factor of the duty-ratio ($d$): The duty-ratio describes the fraction of the total periodically switching cycle where the transistor element is switched on.
\begin{equation}
	V_o=V_i \cdot d
	\label{eq:buck-duty}
\end{equation}

\tikzexternalenable
\tikzexternaldisable

The duty-ratio factor of the input/output voltage relation persists with limitation by the switching operation of the transistor element as long as the switching frequency ($f_{sw}$) is fixed and the circuit has a continuous inductor current ($i_L$) \cite{mohanPowerElectronicsConverters1995}. In the time-domain, the switching frequency ($f_{sw}$) may be expressed as the switching time period ($T_p$).

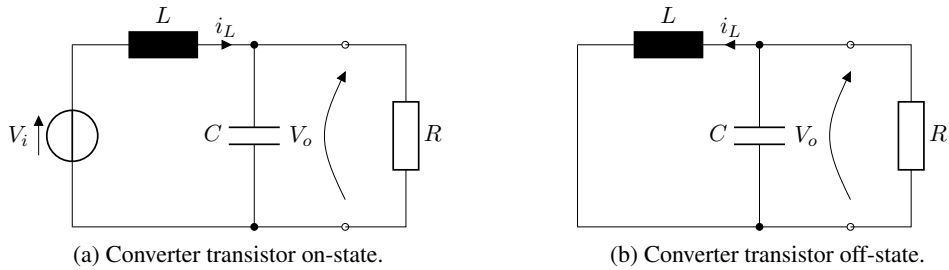
\begin{figure}[htp]
\centering
\subfloat[Converter transistor on-state.\label{fig:buck-states-on}]
	{%
		{\scalebox{0.8}{\begin{circuitikz}
	\draw(0,0) to[european voltage source,v_<=$V_i$] ++(0,-3) coordinate(dc-gnd); %node[](dc-gnd){};
	\draw (0,0) to[short] ++(0,0) coordinate(mid-top) to[L,l=$L$,invert,i=$i_L$] ++(3,0) coordinate(out-m-top) to[C,l_=$C$] ++(0,-3) coordinate(out-m-bot) to[short] ++(-3,0) coordinate(mid-bot) to[short] (dc-gnd);
	\draw (out-m-top) to[short,*-o] ++(1.5,0) coordinate(out-top);
	\draw (out-m-bot) to[short,*-o] ++(1.5,0) coordinate(out-bot);

	\draw (out-top) to[open,v<=$V_o$] (out-bot);
	\draw (out-top) to[short] ++(1,0) to[R,l=$R$] ++(0,-3) to[short] (out-bot);
	
\end{circuitikz}%
}
	}%
}
%\hfill
\hspace*{4em}
\subfloat[Converter transistor off-state.\label{fig:buck-states-off}]
	{%
		{\scalebox{0.8}{\begin{circuitikz}
	\draw(0,0) to[short] ++(0,-3) coordinate(dc-gnd); %node[](dc-gnd){};
	\draw (0,0) to[short] ++(0,0) coordinate(mid-top) to[L,l=$L$,invert,i^<=$i_L$] ++(3,0) coordinate(out-m-top) to[C,l_=$C$] ++(0,-3) coordinate(out-m-bot) to[short] ++(-3,0) coordinate(mid-bot) to[short] (dc-gnd);
	\draw (out-m-top) to[short,*-o] ++(1.5,0) coordinate(out-top);
	\draw (out-m-bot) to[short,*-o] ++(1.5,0) coordinate(out-bot);

	\draw (out-top) to[open,v<=$V_o$] (out-bot);
	\draw (out-top) to[short] ++(1,0) to[R,l=$R$] ++(0,-3) to[short] (out-bot);
	
\end{circuitikz}%
}%
	}
}
	\caption{The circuit equivalents of \cref{fig:buck} as determined by transistor state.}
	\label{fig:buck-states}
\end{figure}

Without further details of the design consideration, and the introduction of semiconductors, transistor and diodes, the buck converter circuit in \cref{fig:buck} can be in two possible states. This is illustrated in \cref{fig:buck-states-on,fig:buck-states-off} as equivalent circuits for the on- and off-state, respectively. As can be observed from \cref{fig:buck-states-on,fig:buck-states-off}, the semiconductor elements is replaced by their feature in the switching state by ideal properties as either fully closed or opened. The most significant observation is the impact on the inductor element ($L$) based on the switching state. In the on-state (\cref{fig:buck-states-on}) the inductor is charged by the source $V_i$ and a current $i_L$ will flow through the inductor in the direction of the output $V_o$. Due to the opposing force of the inductor an induced voltage drop will be generated between the source $V_i$ and the output $V_o$.
In the off-state (\cref{fig:buck-states-off}) the full voltage potential of the output $V_o$ will act on the inductor, giving a voltage potential difference and an oppsite flow direction of the current $i_L$. The opposing electromotive force of the inductor will generate a voltage potential difference that supports the output  $V_o$ with energy. 

Based on the circuit states, the two differential equations \cite{mohanPowerElectronicsConverters1995} describing the circuit based on the change in the input inductor current ($i_L$) and output voltage ($v_o$) are given in \cref{eq:buck-il} and \cref{eq:buck-vo}:

\begin{subequations}
	\begin{equation}
		\dfrac{di_L}{dt}= \dfrac{V_i}{L} (u) - \dfrac{v_o}{L}
		\label{eq:buck-il}
	\end{equation}
	\begin{equation}
		\dfrac{dv_o}{dt}= \dfrac{1}{C}i_L - \dfrac{1}{RC}v_o
			\label{eq:buck-vo}
	\end{equation}
	\label{eq:buck}
\end{subequations}

As can be seen, the transistor switching states only affects the $\dfrac{V_i}{L}(u)$ expression, where $u$ is the switch state, zero (off) and one (on).

\begin{figure}[H]
	\centering
	\scalebox{0.35}{\includegraphics
		{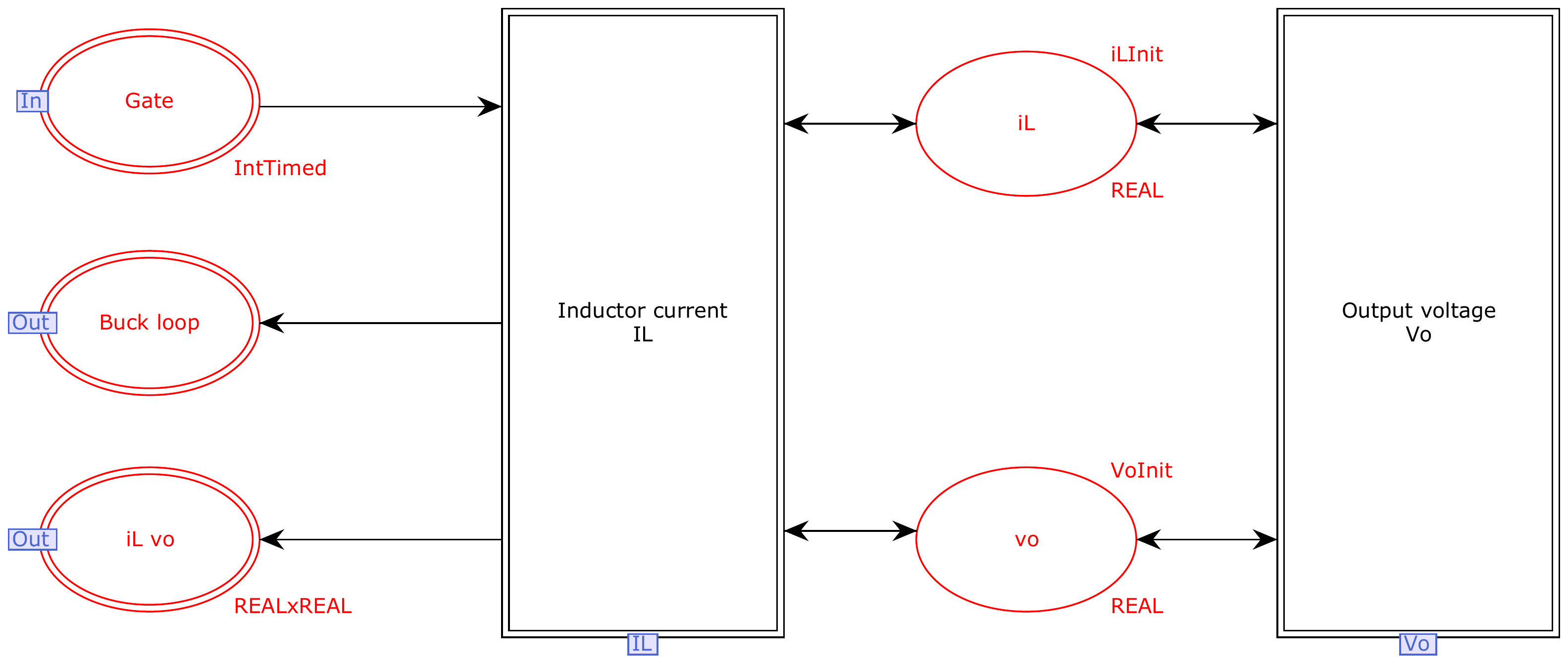}}
	\caption{Submodule of the \cpntxt{Buck Converter} with external interfaces to the controller (left) and calculation of $I_L$, \cref{eq:buck-il} and $V_o$, \cref{eq:buck-vo} represented by substitutions transitions.}
	\label{fig:cpn-buck-model-overview}
\end{figure}

\Cref{fig:cpn-buck-model-overview} shows the buck converter submodule of the \cpntxt{Buck Converter} substitution transition found in the top-level model in \cref{fig:cpn-zynq-buck-full}.
The places \cpntxt{Gate}, \cpntxt{Buck loop}, and \cpntxt{iL vo} are port places of the submodule as can be seen by the \cpntxt{IN}- and \cpntxt{OUT}-tags which determines whether the places is a port place where the submodule can consume tokens or add tokens, respectively. The port places are linked via a port-socket assignment to the accordingly named socket places in \cref{fig:cpn-zynq-buck-full}. This is what makes this module able to exchange tokens with its higher-level module and is used to model the signalling between the buck converter and the controller. It can be seen that the module in \cref{fig:cpn-buck-model-overview} again contains two substitutions transitions \cpntxt{IL} and \cpntxt{Vo}. These submodules associated with these two transitions model the inductor current ($I_L$) and the output voltage ($V_o$) as described by the differential equations from \cref{eq:buck}. The socket places \cpntxt{IL} and \cpntxt{Vo} are used for modelling the interconnected behaviour of the two differential equation in \cref{eq:buck}, i.e., that $v_o$ appears as a term in \cref{eq:buck-il} and that $i_L$ appears as a term in \cref{eq:buck-vo}. The modelling of the inductor current and the output voltage is presented in more detail below.

\subsection{CPN modelling of inductor current (\cpntxt{IL})}
\label{subsec:cpn-il}

\begin{figure}[h]
	\centering
	\scalebox{0.35}{\includegraphics
		%[clip, trim=.5cm 7.5cm 0.5cm .5cm]
		%trim=left botm right top
		%{figs/cpn-buck-model-il-eps-converted-to.pdf}}
	{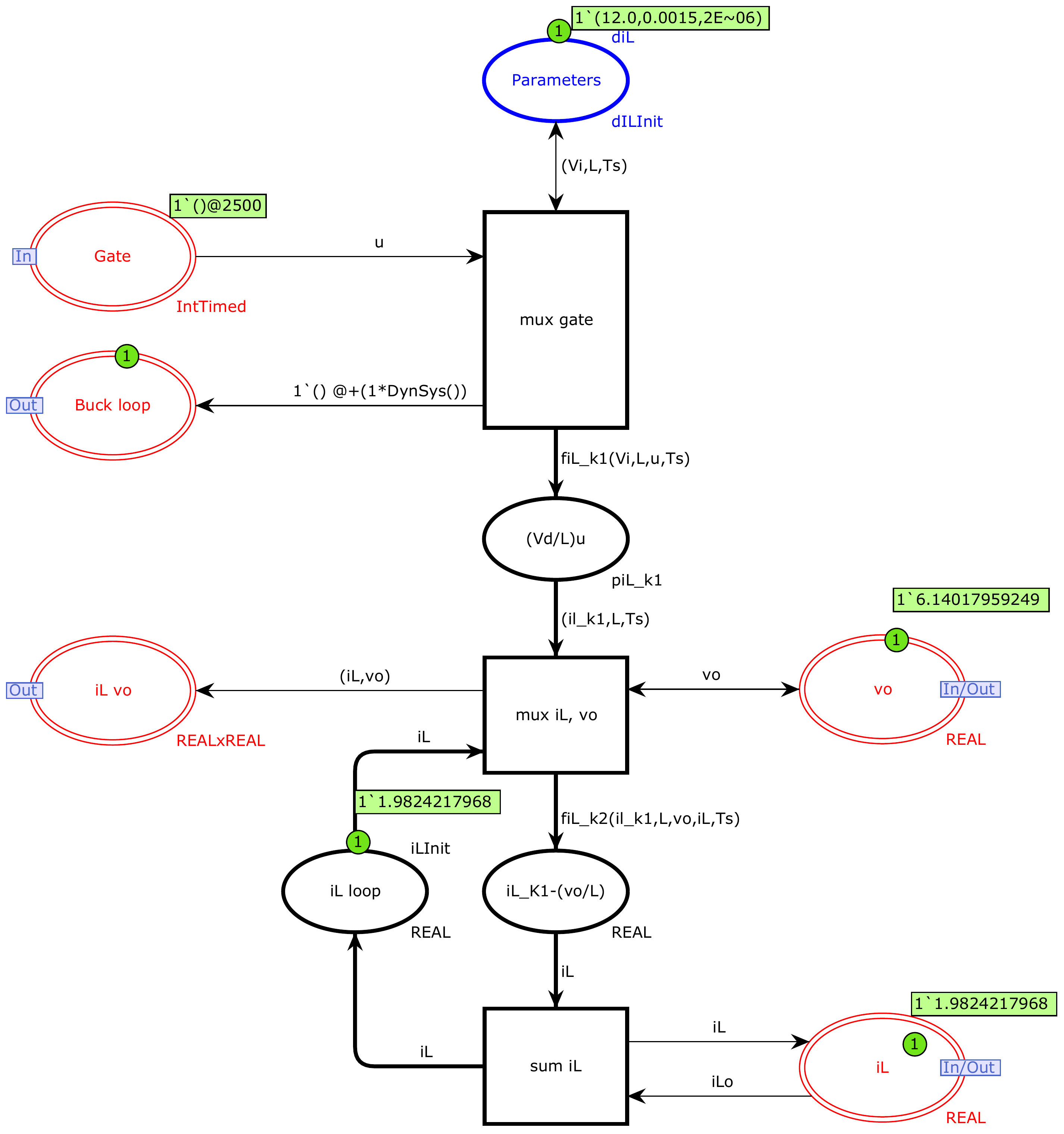}}
\caption{Submodule \cpntxt{IL} estimating the inductor current ($i_L$) based on \cref{eq:buck-discrete-il}.}
\label{fig:cpn-buck-model-il}
\end{figure}

\Cref{fig:cpn-buck-model-il} shows the submodule of the \cpntxt{Inductor Current IL} substitution transition which implements the first part of the integrated behaviour of the buck converter. For illustration purpose an active simulation is selected to show the markings of values and time properties. The port places \cpntxt{iL} and \cpntxt{vo} are both input and output port places of this module as indicated by the \cpntxt{In/Out}-tag and serve as an interface to the submodule \cpntxt{Vo} in order to exchange tokens carrying the values for inductor current ($i_L$) and output voltage ($v_o$).  As described by the differential equations in \cref{eq:buck}, the physical response between the modules is integrated.

The colour sets definitions, variable declarations, and function definitions for the module are shown in \cref{lst:cpn-buck-il}.
The variables and functions are used in the arc expressions in the module to specify the tokens removed from input places and added to output places when an enabled transition occurs. The enabling of a transition in a CPN model is determined by creating a binding of values to the variables occurring in the arc expressions on the surrounding arcs of the transition. A binding represents an enabling mode of the transition. Whether the binding of a transition is enabled in the current marking of the model or not is determined by evaluating the surrounding input arc expression of the transition in the binding. If the multi-set of tokens obtained is a subset of the multi-set of tokens present on the corresponding input place for all input arc expression, then the binding is enabled and may occur. When an enabled binding of a transition occurs it will remove from each input place the multi-set of tokens determined by evaluating the input arc expressions and add tokens to the output places as determined by evaluating the output arc expressions. As an example, consider the \cpntxt{sum iL} transition in \Cref{fig:cpn-buck-model-il}  which has the two variables \smltxt{iL} and \smltxt{iLo} appearing in its surrounding arc expressions. To determine whether this transition is enabled, we need to bind values to the two variables and then evaluate the two input arc expressions to determine whether the required tokens are present on the two input places. If so, then the transition may occur removing the tokens from the input places and added tokens to the two output places as determined by evaluating the output arc expressions.

The purpose of the \cpntxt{IL} module is to estimate the inductor current from \cref{eq:buck-il} based on the switch state and the circuit electrical properties (see \cref{fig:buck}) as it evolves over time.
Due to the requirement of time in the circuit analysis, the description of the inductor current found in \cref{eq:buck-il} must be adapted for discrete-event numerical simulation. As introduced at the end of \cref{sec:cpn-implementation-cps}, the modelling of time includes the iteration constant \smltxt{k} and the real-time conversion constant \smltxt{Ts} in the discretisation.
Based on this discretisation, we obtain the description expressed by \cref{eq:buck-discrete-il} that is implemented by the \cpntxt{IL} module in \cref{fig:cpn-buck-model-il}.
The description is also dependent on the switching element $u$ with either a dynamic or fixed switching frequency ($f_{sw}$) covering the corresponding switching period ($T_p$).

\begin{equation}
	i_L ((k+1)T_s)=\overbrace{T_s \left[\overbrace{\dfrac{V_i}{L} (u)(k T_s)}^{\text{\smltxt{fiL\_k1}}} - \dfrac{v_o}{L}(k T_s) \right]}^{\text{\smltxt{fiL\_k2}}}
	\label{eq:buck-discrete-il}
\end{equation}
As may be observed, the modelling approach as represented by the submodule in \cref{fig:cpn-buck-model-il} closely follows the buck converter description and behaviour as represented via the differential equations.
This demonstrates how we have targeted a transparent modelling approach where the detailed modelling of the components and the steps involved in their execution are fully available for inspection and validation. This is in contrast to the black-box approach adopted by Matlab/Simulink.

The buck converter contains physical components each of which has a value, such as capacitance ($C$), inductance ($L$) and load resistanse ($R$). In addition, the considered operation of the converter is with a fixed value for the input voltage ($V_i$). The value of the components and input voltage forms the constant parameters defined at the place \cpntxt{Parameters} and holds a double-headed arc to the transition \cpntxt{mux gate}. The double-headed arc implies that the place \cpntxt{Parameters} act both as input/output and will hold the defined constant values as the place markings for all transitions.
The occurrence of the \cpntxt{mux gate} transition is triggered based on the received timed controller signal to the gate activation ($u$) from the place \cpntxt{Gate}. The gate activation ($u$) high and low state is represented by 1 and 0, respectively. By adding the constant parameters from the place \cpntxt{Parameters} all variables to enable the firing of the transition \cpntxt{mux gate} exists and the function \smltxt{fiL\_k1} on the arc expression estimates the first part of \cref{eq:buck-discrete-il}.
After each firing of the transition \cpntxt{mux gate} the model timed marking \cpntxt{@} is added and exchanged with the controller through the place \cpntxt{Buck loop}. This means that the model represent and differentiate the behaviour of the converter as an interconnected electrical circuit with simultaneous impact and the timed properties.

%\vspace*{0.5em}
\begin{lstlisting}[caption={Colour sets and functions used in the \cpntxt{iL} submodule.},label={lst:cpn-buck-il}]
	colset IntTimed  = int timed; (* Timed gate control signal *)

	colset Vi        = real; (* Input voltage *)
	colset iL        = real; (* Inductor current *)
	colset iLo       = real; (* Inductor current previous step *)
	colset L         = real; (* Inductance *)
	colset u         = int timed; (* Transistor gate signal *)
	colset Ts        = real; (* Time step *)

	(* Constant values and estimation for the inductor currrent*)
	colset dILInit   = product Vi*L*Ts;
	colset piL_k1    = product iL_k1*L*Ts;

	colset REALxREAL = product iL*vo;

	(* Related to Eq. (3) *)
	fun fiL_k1(Vi,L,u,Ts)        = (if u=1 then (Vi/L)*Ts else	0.0,L,Ts);
	fun fiL_k2(il_k1,L,vo,iL,Ts) = (iL+(il_k1-(vo/L)*Ts));
\end{lstlisting}

The second part of \cref{eq:buck-discrete-il} is estimated using the \smltxt{fiL\_k2} function as part of the firing of the \cpntxt{mux Il,vo} transition and summation of the resulting inductor current ($i_L$) from the first part. \cpntxt{mux Il,vo} is dependant of the output voltage \cpntxt{vo} found in the submodule \cpntxt{Vo} and represent the second part of the differential equation describing the converter (\cref{eq:buck-vo}).
The arc expression from transition \cpntxt{mux gate} to place \cpntxt{Buck loop} is used to signal an acknowledgment which confirms when an iteration of the buck converter model has been executed and thereby allow the next token from the controller via the \cpntxt{Gate} place. Hence, the \cpntxt{Buck loop} interface place is purely a modelling artefact for executing the CPN model correctly wrt. time and does not have a physical counterpart in the real system.

\subsection{CPN modelling of output voltage (\cpntxt{Vo})}

\begin{figure}[h]
	\centering
	\scalebox{0.35}{\includegraphics
		%[clip, trim=.5cm 7.5cm 0.5cm .5cm]
		%trim=left botm right top
		%{figs/cpn-buck-model-vo-eps-converted-to.pdf}}
	{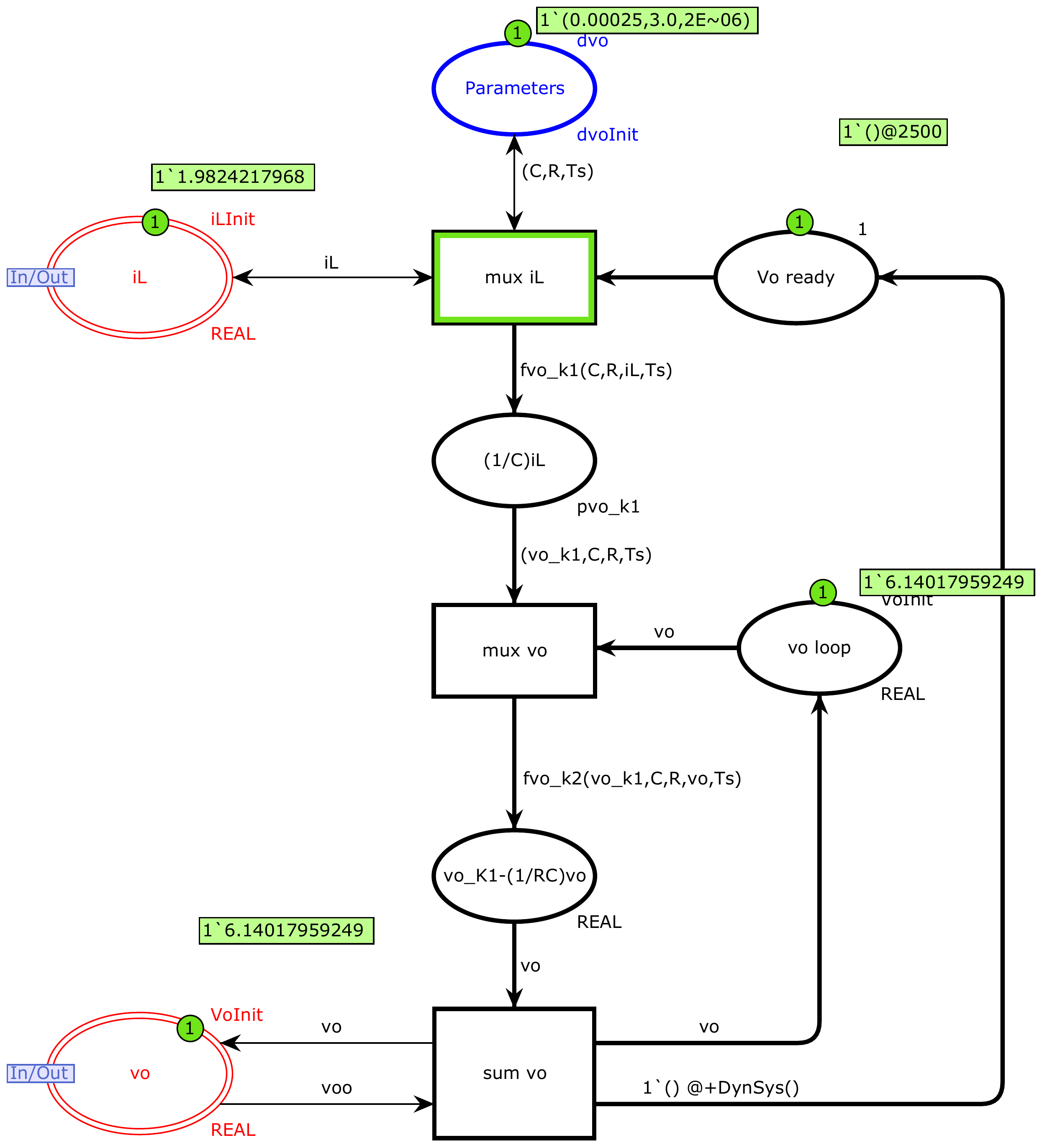}}
\caption{Submodule \cpntxt{Vo} for estimating the output voltage ($v_o$) based on \cref{eq:buck-discrete-vo}.}
\label{fig:cpn-buck-model-vo}
\end{figure}

\Cref{fig:cpn-buck-model-vo} shows the submodule of the output voltage \cpntxt{Vo} substitution transition implementing the second part of the integrated behaviour of the buck converter. The places \cpntxt{iL} and \cpntxt{vo} act as the interface to the submodule \cpntxt{IL}, and exchange the values for inductor current ($i_L$) and output voltage ($v_o$). The colour set definitions and the functions used in the arc expressions are listed in \cref{lst:cpn-colset-buck-vo}.

The output voltage ($V_o$) is dynamically connected to the inductor current ($i_L$) and the corresponding switch-state ($u$) found in the submodule \cpntxt{iL} described in \cref{subsec:cpn-il}. The buck converter physical component characteristics are defined at the place \smltxt{Parameters} with the load resistor ($R$) and capacitance ($C$) and are implemented as constants with a double-arc to the transition \cpntxt{mux iL}. To enable initialization and firing of the transition \cpntxt{mux iL} the place \cpntxt{Vo ready} is added. The place \cpntxt{Vo ready} holds an initial marking to enable the first iteration and are later enabled through the timed marking describing the full execution of the converter model.
By adding the constant parameters from the place \cpntxt{Parameters} all variables to enable the firing of the transition \cpntxt{mux iL} exists and the function \smltxt{fvo\_k1} on the arc expression estimates the first part of \cref{eq:buck-discrete-vo}.
At the transition \cpntxt{mux vo} the previous voltage value $v_o$ is added trough the place \cpntxt{vo loop} to estimate the second part of \cref{eq:buck-discrete-vo} trough the arc expression \smltxt{fvo\_k2}.
The description found in \cref{eq:buck-vo} is extended to include time properties of the discrete-event model using the same discretisation approach as for the inductor current ($i_L$).

\begin{equation}
	v_{o}((k+1) T_s)= \overbrace{T_s\left[\overbrace{\dfrac{1}{C}i_L(k T_s)}^{\text{\smltxt{fvo\_k1}}} - \dfrac{1}{RC}v_o (k T_s) \right]}^{\text{\smltxt{fvo\_k2}}}
	\label{eq:buck-discrete-vo}
\end{equation}

\begin{lstlisting}[caption={Definition of the color set related to substitution transition \cpntxt{Vo}.},label={lst:cpn-colset-buck-vo}]
colset C        = real; (* Capacitance *)
colset R        = real; (* Resistance *)
colset vo       = real; (* Output voltage *)
colset voo      = real; (* Output voltage previous step *)
colset iL       = real; (* Inductor current *)
colset Ts       = real; (* Time step *)

(* Constant values and estimation for the output voltage *)
colset dvoInit  = product C*R*Ts;
colset pvo_k1   = product vo_k1*C*R*Ts;

(* Related to Eq. (4) *)
fun fvo_k1(C,R,iL,Ts)       = (((iL/C)*Ts),C,R,Ts);
fun fvo_k2(vo_k1,C,R,vo,Ts) = (vo+(vo_k1-((Ts)*vo/(R*C))));
\end{lstlisting}

\tikzexternaldisable

\section{Controller CPN Modelling}\label{subsec:controller-interface}

\begin{figure}[h]
	\centering
	\scalebox{0.35}{\includegraphics
		{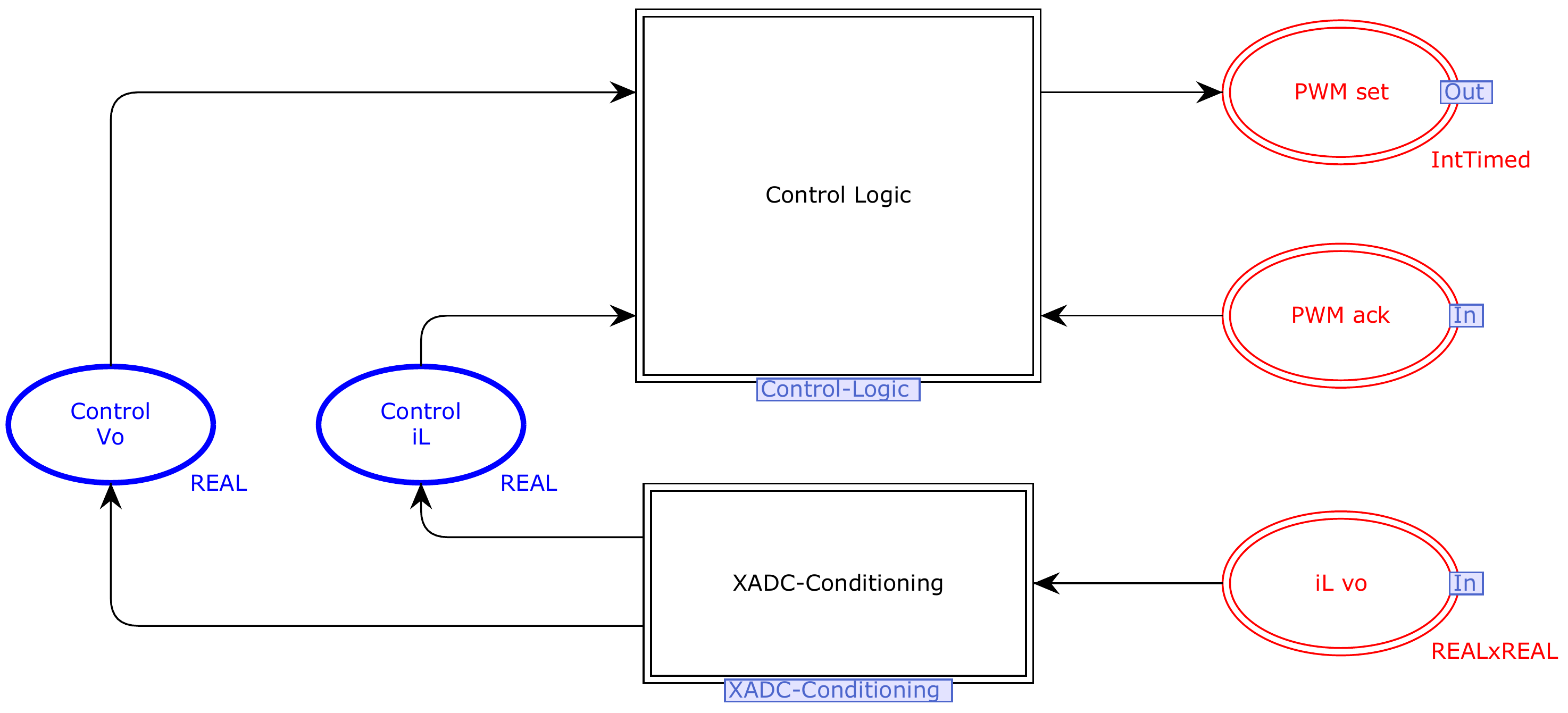}}
	\caption{Overview of \gls{abr-fpga}-\gls{abr-arm} modelling forming the Zynq \gls{abr-soc}.}
	\label{fig:cpn-fpga-arm}
\end{figure}

\Cref{fig:cpn-fpga-arm} shows the \cpntxt{FPGA-ARM Controller} submodule of the \cpntxt{FPGA-ARM Controller} substitution transition in \cref{fig:cpn-zynq-buck-full}. The modelling of the FPGA-ARM controller in \cref{fig:cpn-fpga-arm} is organised into two submodules: the \cpntxt{XADC-Conditioning} and \cpntxt{Control Logic}. The purpose of dividing controller in two parts (\cpntxt{XADC-Conditioning} and \cpntxt{Control Logic}) is to capture the data-flow and connected parsing delays in the Zynq \gls{abr-soc} from sensing of input from the buck converter through execution of the control logic to the eventual effect on the controller output.
The two socket places \cpntxt{Control Vo} and \cpntxt{Control iL} in \cref{fig:cpn-fpga-arm} are used for modelling the interconnection of the two sub-modules. An advantage of having the two dedicated control tasks explicitly separated is that we can easily adopt more complex control logic into the model. The modelling of the control tasks is described in detail below.

\subsection{CPN modelling of the \textmd{\cpntxt{XADC conditioning}}}
\label{subsubsec:xadc-conditioning}

\begin{figure}[h]
	\centering
	\scalebox{0.35}{\includegraphics
		{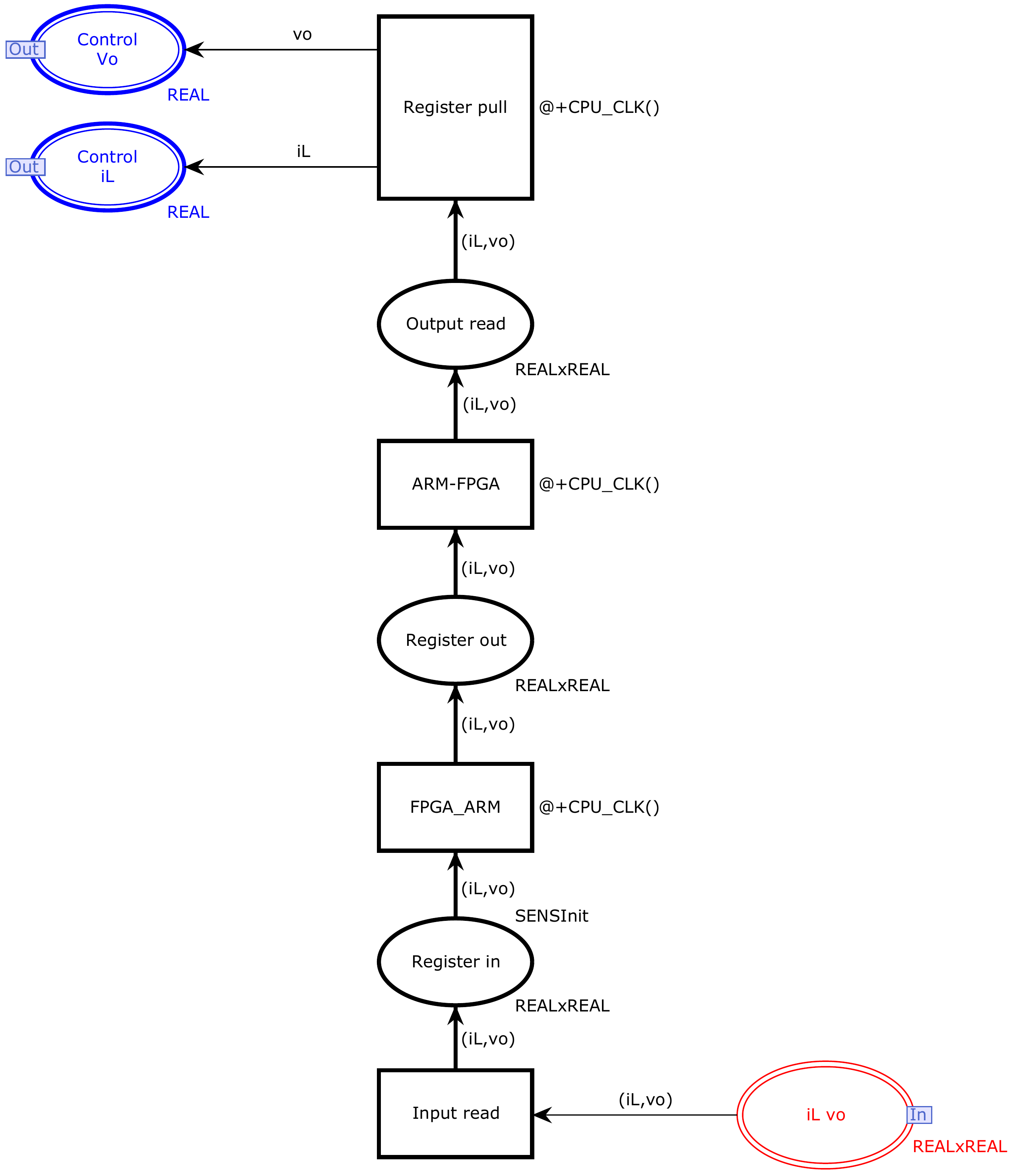}}
	\caption{The \cpntxt{XADC-Conditioning} submodule modelling of FPGA and the ARM interface}
	\label{fig:cpn-xadc-conditioning}
\end{figure}

\Cref{fig:cpn-xadc-conditioning} shows the submodule of the controller signal conditioning circuit associated with the \cpntxt{XADC-Conditioning} substitution transition from \cref{fig:cpn-fpga-arm}. It implements the required steps from parsing the signals from the input, through the \glsxtrfull{abr-adc}, and the preparation of the readings for use in control tasks. The input-pins are allocated to the \gls{abr-fpga} and needs to be transferred to the integrated \gls{abr-adc} for signal conditioning in the \gls{abr-arm}.

Each of the steps, moving the data between registers, requires \gls{abr-fpga} and \gls{abr-arm} processor execution implying a time delay. This is modelled by the aspect of
\smltxt{@+CPU\_CLK()} time inscriptions placed next to the transitions. There is no time delay associated with the \cpntxt{Input read} as it does not involve any CPU processing for the sensed input on place \cpntxt{iL vo} to be available via the associated \cpntxt{Register in}.

\clearpage
\subsection{CPN modelling of the \cpntxt{Control Logic}}

\begin{figure}[h]
	\centering
	\scalebox{0.35}{\includegraphics
		{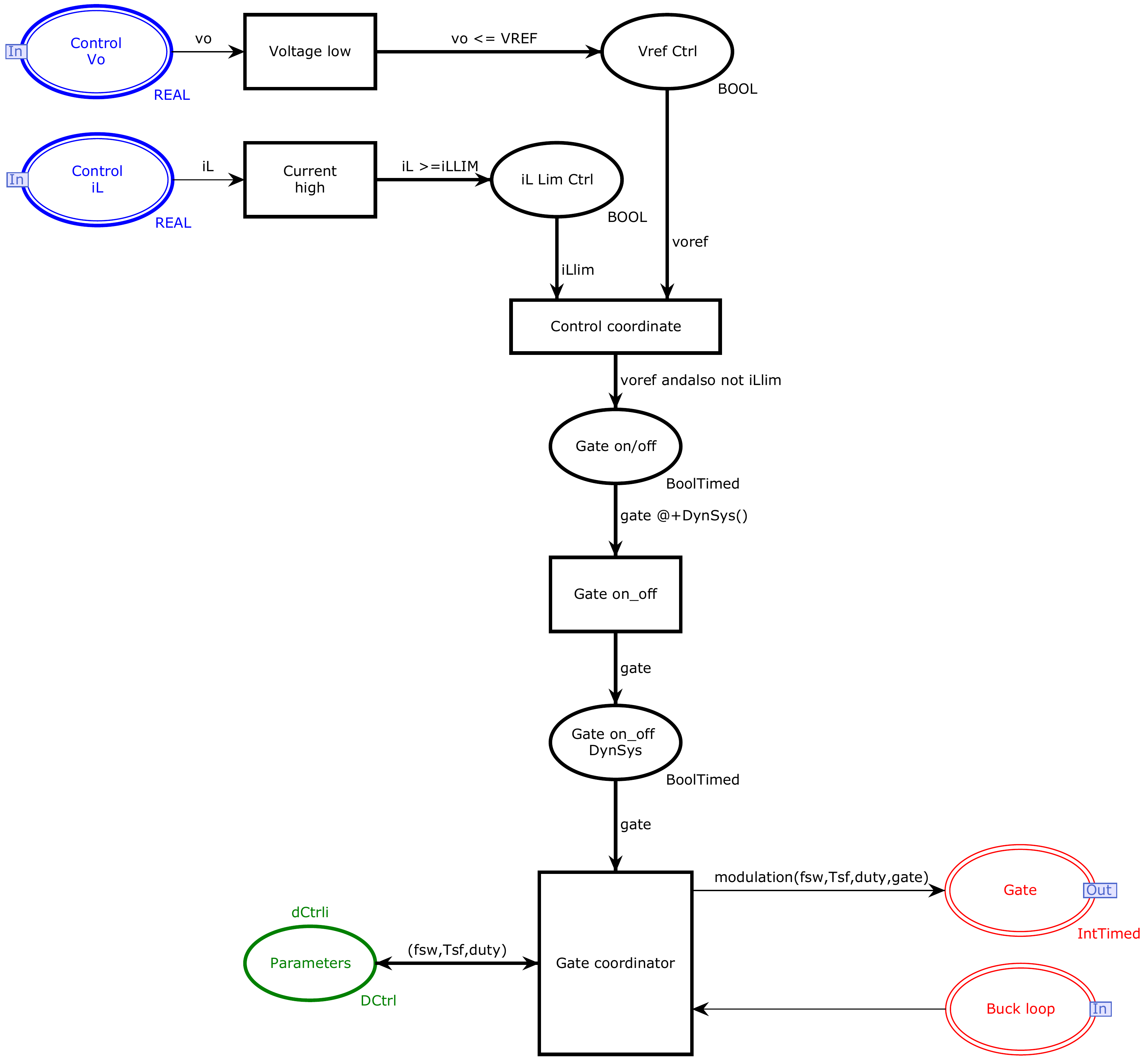}}
	\caption{The \cpntxt{Control Logic} submodule with open-loop and closed-loop controls.}
	\label{fig:cpn-fpga-controller}
\end{figure}

\Cref{fig:cpn-fpga-controller} shows the submodule of the \cpntxt{Controller Logic} substitution transition.
The associated colour set and control functions are defined in \cref{lst:cpn-control-logic}. It models the execution of the control logic in the \gls{abr-fpga} and the setting of the control output to the buck converter through the \cpntxt{Control logic} substitution transition in \cref{fig:cpn-fpga-arm}.
The controller uses the power of the FPGA by executing all logic concurrently based on a single read from the bulk register with the input from the buck converter output voltage, \smltxt{vo}, and inductor current, \smltxt{iL} from the \cpntxt{XADC-Conditioning}. The time control is therefore determined by the \cpntxt{XADC-Conditioning} module, adding a \smltxt{@+CPU\_CLK()} for the pull request of the analog sensing values to the controller.

\noindent
Three modes of control are represented by the model of the controller logic:
\begin{enumerate}
	\item The duty ratio control which ensures a gate switching at given duty ratio ($d$) of the switching frequency, $f_{sw}$. With this operation, the ideal properties of the converter input/output voltage relation is shown in \cref{eq:buck-duty}.
	\vspace{2em}
	\item Voltage reference control ensures that the gate is attempting to compensate a voltage deviation based on the sensing of the output voltage, \smltxt{vo}, to the voltage reference \smltxt{VREF} as closed-loop control.
	\item Inductor current protection ensures that the gate is not overloading the inductor and converter by sensing the inductor current, \smltxt{iL}, and are blocking the gate if the threshold current \smltxt{iLLIM} is exceeded.
\end{enumerate}
The gate signal ($u$) is modulated at the output stage between the controller and buck converter with the function \smltxt{modulation}.
Through the modulation function, the predefined switching frequency \smltxt{fsw} and time-step discretisation factor \smltxt{Tsf} is aligned with the real-time properties of time-step \smltxt{Ts}. The predefined parameters are defined at the place \cpntxt{Parameters}.
As the duty ratio, \smltxt{duty}, is a fraction of the switching time periode \smltxt{Tp} it also implies that the duty ratio is a fraction of the number of the time-step discretisation factor \smltxt{Tsf}.
In the \smltxt{if}-statements, the relative duty ratio to the switching period \smltxt{Tp} is compared to the real-time execution of the model. In addition, the gate output considers the \smltxt{gate}-conditions from \smltxt{VREF} and \smltxt{iLLIM}.

\vspace*{1em}
\begin{lstlisting}[caption={Definition of the colour set related to substitution transition \cpntxt{Control logic}.},label={lst:cpn-control-logic}]
colset DCtrl = product fsw * Tsf * duty;

fun modulation(fsw,Tsf,duty,gate) =
	let
		val time_stamp       = time()
		val time_stamp_real  = ((IntInfToReal 2 time_stamp)*100.0)
		val Tp               = 1.0/fsw
		val Ts               = Tp/Tsf

		(* Sets the duty ratio relativ to the simulator timesteps *)
		val relative_duty    = Tsf*duty

		(* Defines the count of cycles, Tp, relative to simulator timesteps *)
		val TpCount          = Real.fromInt(round (((IntInfToReal 0 (time()))/Tsf)-0.50))
	in
		if TpCount<1.0 (* As initial point the Tp counter is ignored*)
		then
			if relative_duty > time_stamp_real andalso gate then 1 else 0
		else
			if relative_duty > (time_stamp_real-(Tsf*TpCount)) andalso gate then 1 else 0
	end;
\end{lstlisting}

\section{Simulink Model of the Buck Converter}\label{sec-simulink}

For validation and comparison, we have implemented the buck converter using the standard component library of Simscape (v 5.1) and Simscape Electrical (v 7.5) under Matlab/Simulink R2021a.

The topology of \cref{fig:buck} is adopted, adding voltage source, switching elements, passive components and measurements.
All Simulink model components have simplified real characteristics which is defined by modifying the parameters of each individual part. In simulation, the discrete solver is selected to catch transients from the switching events consistently and the results are gathered from the scope measurement for output voltage ($v_o$) and inductor current ($i_L$). The model has simple operation of \gls{abr-pwm} fixed duty ratio control with open-loop feedback, controlled by the signal generator connected to the transistor gate.

As may be observed, the Simulink model is implemented with ease based on standard components from a library. However, the detailed modelling of each component is not openly accessible and how Matlab/Simulink implements the circuit approximation is not transparent in an approach to evaluate the model step execution.
This gives the CPN Tools model a benefit by its transparent representation and the added value for the user can be examined by comparing the results from the Matlab/Simulink model.
In addition, a Matlab/Simulink model that forms the physical controller hardware with its control dynamics is not found publicly available nor methods on how to implement such a system.

\begin{figure}[ht]
	\centering
	\scalebox{0.55}{\includegraphics
		[clip, trim=2.8cm 5.3cm 2.8cm 5.3cm]
		%trim=left botm right top
		{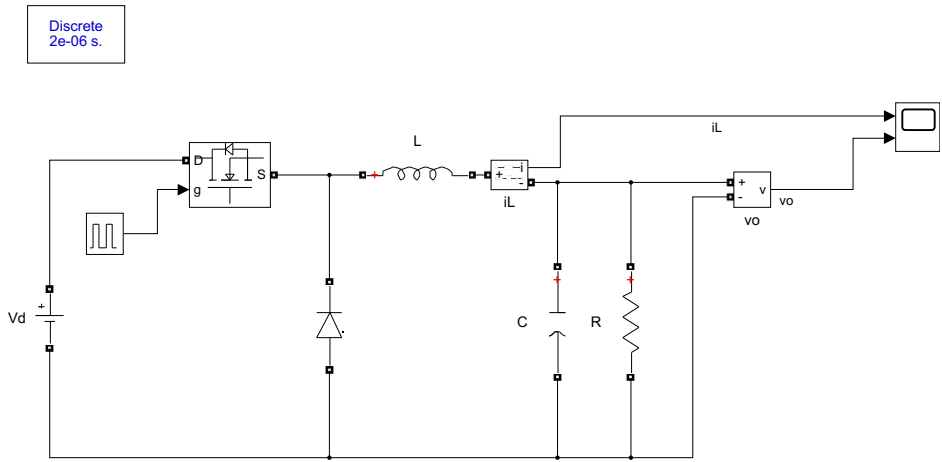}}
	\caption{Matlab/Simulink model of the buck converter without controller}
	\label{fig:matlab-buck-converter}
\end{figure}

\tikzexternalenable
\section{Model Validation and Experimental Results}\label{sec-results}

Our experimental model validation is organised in two parts. In the fist part, the open-loop response is simulated in both CPN Tools and Matlab/Simulink. The goal of this simulation is to ensure the correct implementation of the developed CPN model of the buck converter by comparing the results in a controlled environment. The controlled measure is the scenario of fixed duty ratio ($d$) which draws the attention to the buck converter dynamical performance. In both models, we collect the measured dataset for inductor current ($i_L$), output voltage ($v_o$), and the gate signal ($u$). In CPN Tools, we obtain the simulation results using simulation monitors associated with the places corresponding to the measured values, and store the observed values in external data-files for post-processing and visualisation.

In the second part, the simulations are performed in CPN Tools and forms the integrated \gls{abr-cps} by interfacing the controller and buck converter. The initial closed-loop response is shown based on the \gls{abr-fpga} controller functions and the identified time delays added in the interfacing loops. The results are compared to the hardware setup of the buck converter and controller in an open-loop configuration.
The results are found in \cref{subsec:result-cpn-matlab,subsec:result-cpn-closed-open}, respectively and are based on the component and control parameters given in \cref{tab:results-param}.

\begin{table}
	\caption{Model parameters used in the simulation and physical experiments}
	\label{tab:results-param}
	\centering
	\begin{tabularx}{0.9\textwidth}{ccc>{\raggedright\arraybackslash}X}
		\hline
		Section & Symbol &  Value &  Description \\
		\hline
		\multirow{4}{*}{\ref{subsec:result-cpn-matlab} and \ref{subsec:result-cpn-closed-open}}
		&$f_{sw}$ & \SI{200}{\kilo\hertz} & Switching frequency of the transistor\\
		&$T_s$ & \SI{0.05}{\mu\second} & Discrete time-step \\
		& $V_i$ & \SI{12}{\volt} & Input voltage source \\
		&$L$ & \SI{9.5}{\milli\henry} & Inductor inductance \\
		&$C$ & \SI{20}{\mu\farad} & Capacitor capacitance at the output \\
		&$R$ & \SI{2.4}{\ohm} & Resistor resistance as the load \\
		\hline
		\multirow{ 2}{*}{\ref{subsec:result-cpn-matlab}} && & \\
		&$d$ & \SI{50}{\percent} & Duty ratio of switching frequency\\
		\hline
		\multirow{4}{*}{\ref{subsec:result-cpn-closed-open}}
		&$d$ & \SI{10}{\percent} & Duty ratio of switching frequency\\
		&$V_{REF}$ & \SI{1.2}{\volt} & Output voltage reference\\
		&$I_{LIM}$ & \SI{0.7}{\ampere} & Inductor current limit \\
		\hline
	\end{tabularx}
\end{table}

\subsection{Open-loop response evaluation}\label{subsec:result-cpn-matlab}

\begin{figure}[h]
	\centering
	\tikzsetnextfilename{p-cpn-matlab-duty-compare-physical}
	\subfloat[Output voltage ($v_o$) and inductor current ($i_L$) for the CPN model and the Matlab/Simulink model.\label{fig:openloop-respons-compared-voil}]
	{
	\scalebox{1}{\includegraphics{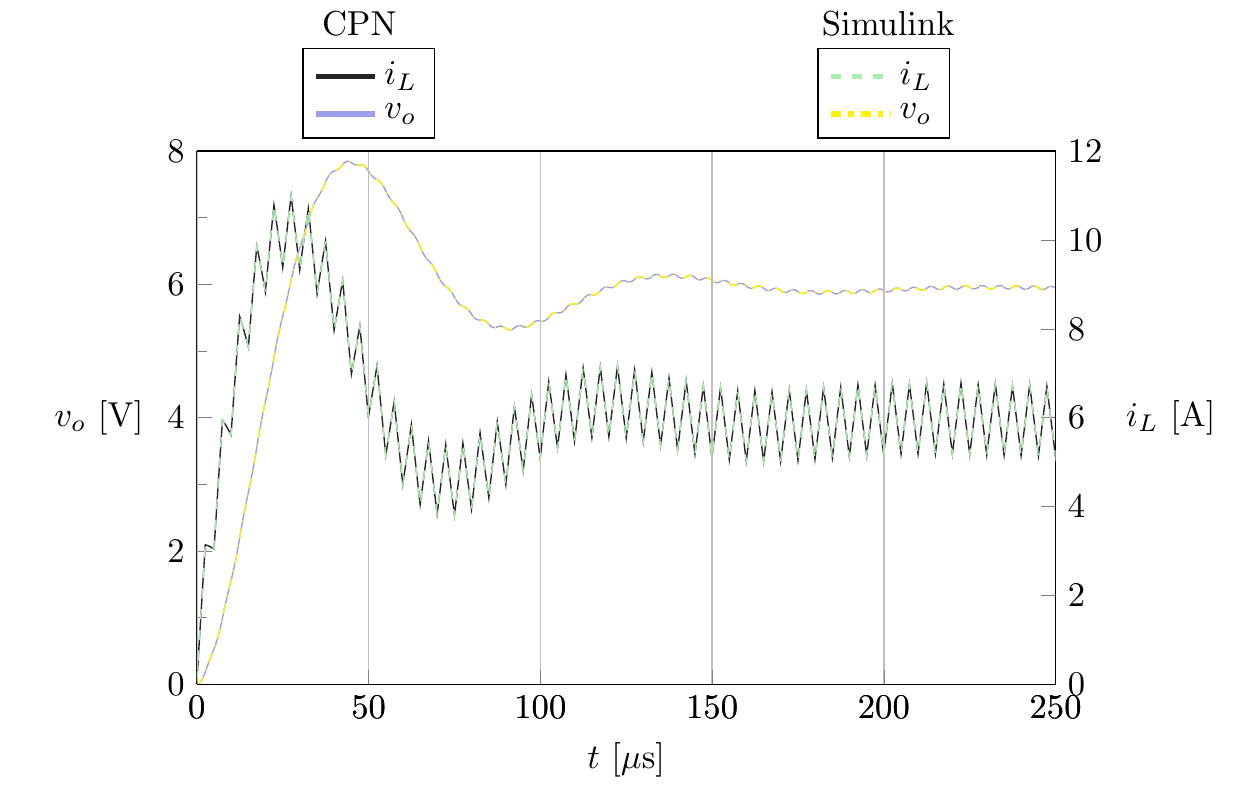}}
	%\subimport{plots}{p-cpn-matlab-duty-compare-physical.tex}
	}
	\\
	\tikzsetnextfilename{p-cpn-matlab-duty-compare-physical-gate}
	\subfloat[	The gate signal operating at duty ratio $d$=\SI{50}{\percent} of the switching frequency, $fsw$=\SI{200}{\kilo\hertz}. The signal is collected from the CPN model.\label{fig:cpn-matlab-openloop-compared-gate}]
	{
	\scalebox{1}{\includegraphics{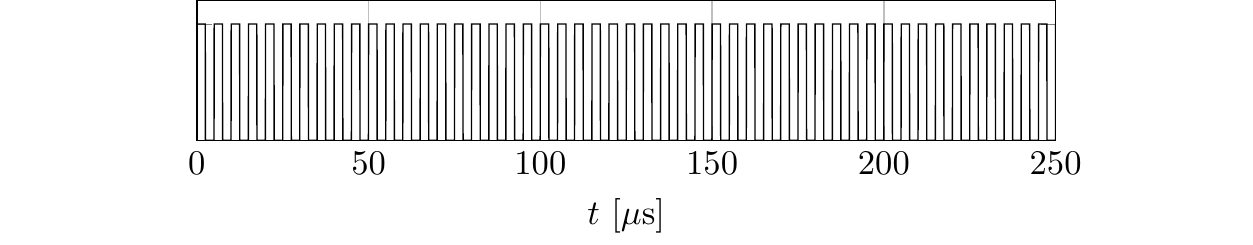}}
	%\subimport{plots}{p-cpn-matlab-duty-compare-physical-gate.tex}
	}
	\caption{Comparing of open-loop response from initial zero state to duty radio $d$=\SI{50}{\percent} with the CPN model and the Matlab/Simulink model.}
\label{fig:cpn-matlab-openloop-compared}
\end{figure}

The simulation results in \cref{fig:cpn-matlab-openloop-compared} shows the initialization and transient response of the buck converter in open-loop configuration at fixed duty ratio ($d$) of the switching frequency, $f_{sw}$=\SI{200}{\kilo\hertz}.
In \cref{fig:openloop-respons-compared-voil} the output voltage ($v_o$) and inductor current ($i_L$) are shown for both the CPN Tools and the Simulink model. The measurements are overlayed, where the CPN Tools measurements are represented with solid lines and the Simulink measurements are dashed.
It can be seen that the transient response of the CPN Tools model correspond perfectly with the Simulink model for both the output voltage ($v_o$) and inductor current ($i_L$).
The inductor properties are clearly interconnected correctly as can be seen from the relation between the gate triggering signal ($u$) in \cref{fig:cpn-matlab-openloop-compared-gate} and the corresponding response on the inductor current ($i_L$) in \cref{fig:openloop-respons-compared-voil}.

\subsection{CPN Tools and physical response evaluation}\label{subsec:result-cpn-closed-open}

\begin{figure}[h]
	\centering
	\tikzsetnextfilename{p-cpn-cl-physical-ol-compare}
	\subfloat[Closed-loop response from initial zero state to a voltage reference set-point of \SI{10}{\percent}. Compared with open-loop response of physical CPS setup.\label{fig:cpn-controller-fpga-voltage-vref}]
	{
		\scalebox{1}{\includegraphics{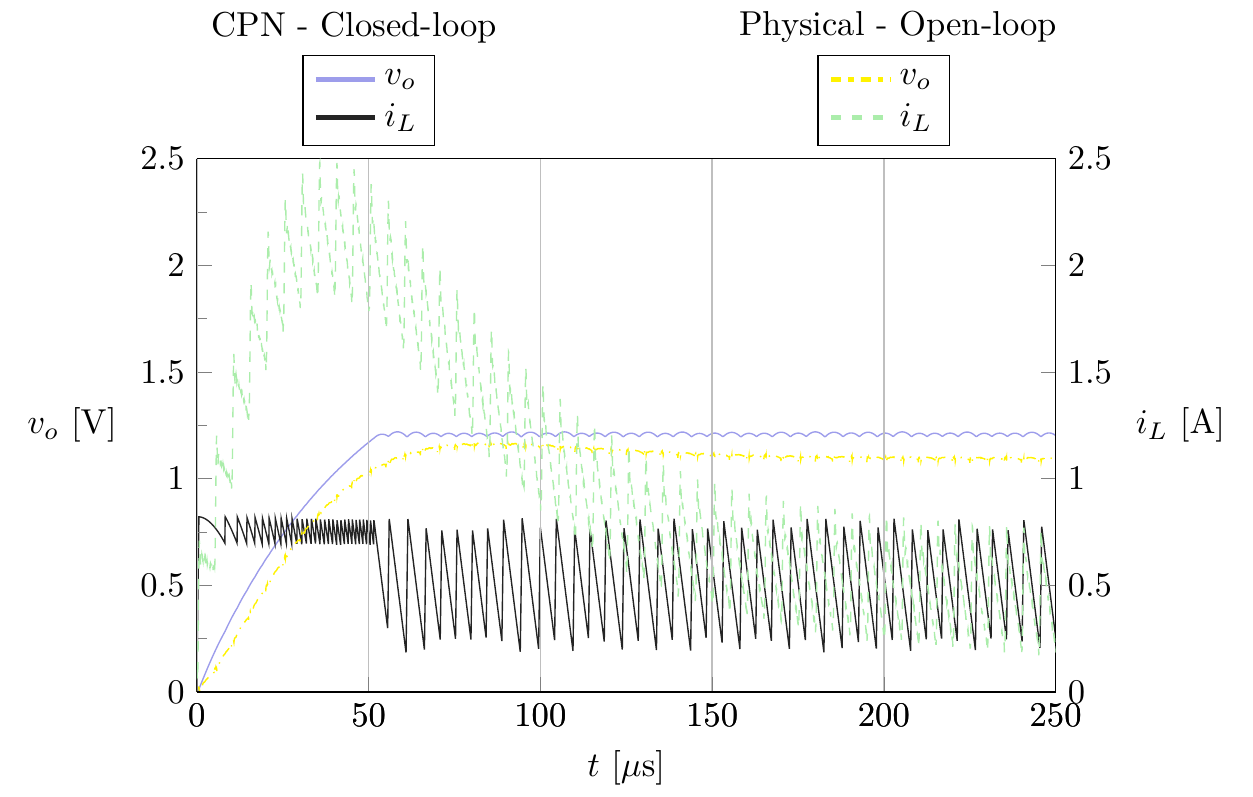}}
		%\subimport{plots}{p-cpn-cl-physical-ol-compare}
	}
	\\
	\tikzsetnextfilename{p-cpn-cl-physical-ol-compare-gate}
	\subfloat[The gate signal operating at dynamic duty based on the closed-loop $V_{\mathit{ref}}$ CPN control.\label{fig:cpn-controller-fpga-voltage-vref-gate}]
	{
		\scalebox{1}{\includegraphics{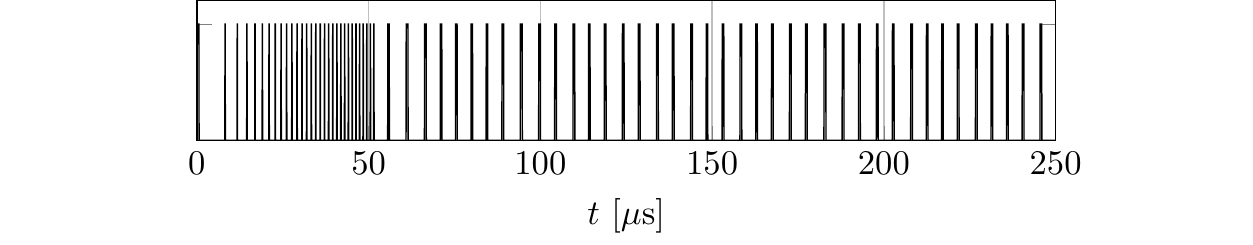}}
		%\subimport{plots}{p-cpn-cl-physical-ol-compare-gate}
	}
	\caption{Comparison of the response from initial zero-state of the closed-loop CPN model, FPGA $V_{ref}$-controller and open-loop control physical setup with duty ratio $d$=\SI{10}{\percent}}
	\label{fig:cpn-controller-fpga-voltage}
\end{figure}

The results in \cref{fig:cpn-controller-fpga-voltage} are comparing the closed-loop control features of the CPN Tools model and the physical measurements of the open-loop buck converter setup (see \cref{fig:physical-buck-zynq-scope-temp}). In CPN Tools, the closed-loop control uses the output voltage ($vo$) measurements and continuously compare against the predefined reference value $V_{REF}$. In addition, the inductor current ($i_L$) protection function is used to limit the maximum current to $I_{LIM}$.
The physical system operates at fixed duty ratio, $d=\SI{10}{\percent}$, of the switching frequency, $f_{sw}$=\SI{200}{\kilo\hertz}.

The transient response of the CPN model with closed-loop controls tends to give an even more rapid output increase, increased steady-state and a more critically damped output voltage ($v_o$) response than the open-loop configuration of the physical system.
When evaluating the simple implementation of the controller this is expected as the gate is switched on as long as the output voltage ($v_o$) is below the set-point value, as a consequence of the control logic for $V_{\mathit{ref}}$ shown in \cref{fig:cpn-fpga-controller}. The additional inductor current protection function is also acting in the initial phase, $t<\SI{50}{\mu\second}$, and limiting ($i_L<I_{LIM}$) which is a favourable feature in a design. An important note with practical limitation for the CPN Tools controller is the high switching frequency observed in the initial phase ($t<\SI{50}{\mu\second}$), which is observed in \cref{fig:cpn-controller-fpga-voltage-vref-gate}. The physical transistor switching speed has some limitation which needs to be considered in converter design and also when considering limitation in control features.

%\tikzexternaldisable

\section{Conclusions and Future Work}\label{sec-sum-conc}

%% summary / conclusions on work don
A power electronics buck converter with associated controller has been modelled
as an integrated system by applying the formalism of Coloured Petri Nets
(CPNs) using the CPN Tools. A key aspect of our CPN model is that it provides full transparency on how both the physical parts and the controller have been modelled. Furthermore, our modelling approach based on CPNs has been developed with a view towards being extendable to more complex converter topologies and systems of power converters.

% summary of experimental resutls
The CPN model of the buck converter was first validated without considering
the controller and feedback loop (in open-loop) with the implemented
Matlab/Simulink model. The results from both models show similar characteristics for both the output voltage ($v_o$) and the inductor current ($i_L$).
The CPN buck converter model was further integrated with a CPN model of the controller to form a fully integrated
CPS. The integrated model shows promising results from operating in closed-loop controls by utilizing the features of the Zynq SoC FPGA controller compared to open-loop response. The results that we have obtained show a rapid voltage increase and good steady-state operation which corresponds to the physical system response and validates the implementation.

% Related work
Petri nets have been widely used for modelling, formal validation, and verification of concurrent systems. In the domain of power electronics converters and embedded systems, \cite{salinasControlDesignStrategy2016} presents a methodology based on low-level Petri nets. The methodology of \cite{salinasControlDesignStrategy2016} is based on associating places with the physical sensing and using transitions occurrences to model the switching in the power electronics elements. A similar approach applying Petri nets was reported in \cite{wisniewskiDynamicPartialReconfiguration2018,wisniewskiDesignVerificationCyberPhysical2019,ortizNovelStrategyDynamic2020}.
There are also examples of extremely high frequency converters of small size, using asynchronous circuits, which applies an interpreter of Petri Nets with \gls{abr-stg} as a framework for the controller logic \cite{sokolovAutomatingDesignAsynchronous2020}. Further the approach of \gls{abr-stg} has been used to implement protection functions related to the transistor switching \cite{liModellingReversionLoss2019}.
Reverse approaches can be found in \cite{beraRelationshipSimulinkPetri2014,mohamedlamineComparisonPWMPetri2017a,alcarazFrameworkBasedMatlab2019}, where the state- and control flow statements in Matlab/Simulink are used to encode Petri net constructs. As discussed in \cite{beraRelationshipSimulinkPetri2014}, Petri nets are well-suited for modelling power electronics systems as they can be represented as a \gls{abr-des} whose state evolution depends on discrete events over time in the form of switching events. Still, as our literature study has shown, the application of Petri nets to power electronics converter systems is limited. Furthermore, the approaches and results reported in the literature have primarily served a design objective by applying Petri nets as a basic framework, and there is a lack of proven track record demonstrating validation capabilities with a Petri net-based approach.

%% Future work
There are several interesting directions to advance the modelling approach developed in this paper. 
The basic component description can be advanced to include unideal characteristics (parasitics effects) and further apply as a general component library. 
In addition, the convert system could be implemented as separate components of linearized models to form more complex power electronics converters and option to simulate the dynamics with systems of converters.
This may allow introduction of more advanced classical analytical tools to assess system stability and frequency response.
The currently implemented closed-loop control is simple, and many improvements are required to ensure safe operation. The potential for exploring such more advanced control schemes is easily achievable with our CPN modelling approach of the buck converter system. Another direction is in the detailed modelling of components and integration to benefit
from the formalism in system design of the CPS and to conduct a more comprehensive validation against a real hardware implementation. With the
higher level of details, more advanced and time-critical control algorithms may
be explored and validated to ensure correct operation. As part of the current
research trends and challenges in improving DC-microgrids, complex converter
topologies such as MMC and systems of converters consist of time-critical control
objectives where CPNs may be applicable. Investigating a top-down approach, considering the overall system design and improvements
to the control architecture is also a possible direction for future work. With such an approach, the level of details may be added gradually and better support an iterative design approach. It may serve both in the
initial design phase with model checking and verification of the structural properties
to avoid formal errors and mistakes in prototyping or when implementing
the higher-level control schemes.

\bibliographystyle{fundam}
\bibliography{cpn-collect}

%\input{template-info}

%%%%%%%%%%%%%%%%%%%%%%%%%%%%%%%%%%%%%%%%%%%%%%%%%%%%%%%%%%%%%%%%%%%%%%

\end{document}